\documentclass[nofootinbib,superscriptaddress,preprint]{revtex4}
\usepackage{amsmath, amsthm, amssymb, amsfonts, braket,dcolumn}
\usepackage{color}
\usepackage{graphicx}
\usepackage{epsfig}

\usepackage[normalem]{ulem}  

\renewcommand\sout{\bgroup\color[rgb]{1,0,0} \ULdepth=-.5ex \ULset}

\newcommand\soutnh{\bgroup\color[rgb] {1.0,0.5,0.35} \ULdepth=-.5ex \ULset}

\begin{document}
\title{Quasi-elastic electron scattering with KIDS nuclear energy density functional}
\author{Hana Gil}
\affiliation{Center for Extreme Nuclear Matter, Korea University, Seoul 02841, Korea}
\author{Chang Ho Hyun}
\email{hch@daegu.ac.kr}
\affiliation{Department of Physics Education, Daegu University, Gyeongsan 38453, Korea}
\author{Kyungsik Kim}
\email{kyungsik@kau.ac.kr}
\affiliation{School of Liberal Arts and Science, Korea Aerospace University, Goyang 10540, Korea}
\date{\today}
\begin{abstract}

Isoscalar and isovector effective masses of the nucleon in nuclear medium are explored in the
quasi-elastic electron scattering off nuclei with KIDS (Korea-IBS-Daegu-SKKU) density functional model.
Effective masses are varied in the range $(0.7 - 1.0) M$ where $M$ is the mass of the nucleon in free space.
Parameters in the KIDS functional are adjusted to nuclear matter equation of state, energy and radius of selected nuclei,
and effective mass of nucleons.
Hartree-Fock equation is solved to obtain the wave functions of the nucleon in target nuclei,
and they are plugged in the calculation of electron-nucleus scattering cross sections at the energies of
incident electrons 300 MeV -- 2.5 GeV.
Theoretical prediction agrees well with measurement.
Dependence on the effective mass is evident: cross section tends to increase with small isoscalar effective masses.
However, effect of isovector effective mass is negligible.
Spectroscopic factors are estimated for the protons in the outermost shells of $^{16}$O, $^{40}$Ca, and $^{208}$Pb.
Results are consistent with the values in the literature.

\end{abstract}
\maketitle
\section{Introduction}

Quasi-elastic electron-nucleus scatterings are acknowledged as a useful tool 
for studying the structure of nuclei and the change of properties of nucleon in nuclear medium.
In the conventional approaches, theoretical models are calibrated to accurate data of nuclei,
so the basic properties of nuclei such as binding energy and charge radii agree to data with differences less than 1\% in most models.
However, model dependence becomes manifest in the structural details.
Representative example is the single particle level: SLy4 \cite{sly4} and UNEDF \cite{unedf0} models 
which are well-known non-relativistic models show discriminate results of the single particle levels in light and heavy nuclei.
Since the protons in specific levels
contribute to the scattering with electrons, model dependence could have effect and appear in the result of cross section.

Effective mass of the nucleon in nuclear medium has long been a controversial issue in the physics of nuclear structure and dense nuclear matter.
It is related to diverse phenomena such as the restoration of chiral symmetry at high densities, density distribution in nuclei,
and single particle energy levels.
Even though many experimental and theoretical efforts have been accumulated, value of the effective mass is not constrained precisely:
generally accepted range is $(0.7 - 1.0) M$ where $M$ is the nucleon mass in free space.

Exclusive $(e,e'p)$ reaction furnishes a tool
useful for studying single particle properties of target nucleus.
It is well known that the process is sensitive to individual orbits and energy levels.
The response functions that give valuable information about the nucleons inside a nucleus can be
extracted from a given orbit with respect to transfers of momenum and energy by comparing with experimental data.
A scale factor called spectroscopic factor (SF) should be determined when theoretical results are compared with data.
Spectroscopic factor contains information about the probability to occupy a given orbit,
so the SF takes a value between 0 and 1 in a nuclear model.
Since the SF represents the structural characteristic of a nuclei, it is useful to estimate the value of the SF for testing a nuclear model.

In this work, we investigate the sensitivity of effective mass to the structure and scattering of nuclei with electrons.
In the KIDS (Korea-IBS-Daegu-SKKU) density functional model, 
one can fix effective masses to specific values without altering the nuclear matter equation of state and static properties of nuclei \cite{kidsnuclei1}.
Effective masses are considered in the range $(0.7 - 1.0) M$.
After model parameters are determined from the nuclear matter properties, nuclear data, and assumed effective masses,
non-relativistic wave equation is solved, and the resulting wave function is transformed to a form adaptable in the relativistic formalism.
Scattering cross sections with protons in the outermost shells are calculated and compared with data.
To include the final state interaction, the wave functions of the continuum nucleon are obtained from the relativistic optical model \cite{clark} for a knocked-out proton solving the Dirac equation.
The incident and outgoing electron Coulomb distortions are treated by the same method as the Ohio group \cite{kim}.

In the result, we find that density distribution around surface shows rare dependence on the effective mass.
However, the dependence becomes evident in the core of heavy nuclei.
Single particle levels also depend on the effective mass.
In the light nuclei, small effective mass gives better agreement to data,
but for $^{208}$Pb, data are reproduced well with effective mass close to the free space mass.
Cross section is in good agreement with experiment.
Dependence on the effective mass is manifest: smaller effective mass tends to give larger cross sections.
Spectroscopic factors are extracted by adjusting the cross sections calculated
from theory to data.
Results are comparable with values of the literature.

Remaining part of the manuscript presents the following contents.
In Section II, we introduce the basic formalism.
Results of the 
density distributions, single particle levels, 
and the cross sections are displayed and discussed in Section III.
In Section IV, we summarize the work.

\section{Formalism}

\subsection{Relativistic wave function from a non-relativisitc nuclear model}

Electrons in the process we consider are highly relativistic, so the available code is constructed in the relativistic
formalism in which both electrons and nucleons are treated relativistically.
Many and well-known nuclear structure models such as Skyrme force models and Gogny force models are
formulated on the non-relativistic ground.
KIDS model is also based on the non-relativistic phenomenology, so it is necessary to bridge the two approaches. 

In the relativistic formalism, single particle wave function with angular momentum $\mathbf{J}^2$ and $J_z$, 
and in good quantum state of parity and time reversal takes the form
\begin{eqnarray}
\Psi(\mathbf{r}) =
\left(
\begin{array}{c}
f(r) \chi^\mu_\kappa (\hat{\mathbf{r}}) \\
i g(r) \chi^\mu_{-\kappa} (\hat{\mathbf{r}})
\end{array}
\right).
\end{eqnarray}
Orbital and spin states are represented by 
\begin{equation}
\chi^\mu_\kappa (\hat{\mathbf{r}}) = \sum_{m, s} \left< l m \left. \frac{1}{2} s \right| j \mu \right> Y_{lm}(\hat{\mathbf{r}}) \chi_s,
\end{equation}
where $\chi_s$ and $Y_{lm}$ are the Pauli spinor and the spherical harmonics, respectively.
$\kappa$ is the eigenvalue of the operator $(\mathbf{\sigma} \cdot \mathbf{L} +1)$ given by
\begin{equation}
\kappa = \left(
\begin{array}{c l}
l & ;\, j = l - \frac{1}{2}, \\
-l-1 & ;\, j = l + \frac{1}{2}.
\end{array}
\right.
\end{equation}
Solving the non-relativistic wave equation
\begin{equation}
\frac{d^2 F}{dr^2} + \frac{2}{r}\frac{dF}{dr} - \frac{\kappa (\kappa+1)}{r^2}F
- 2M \left[ V_{\rm cen}(r) + V_{\rm SO}(r) (-\kappa-1) \right] F + p^2 F = 0,
\label{eq:Fr}
\end{equation}
one can determine $F(r)$.
Radial functions $f(r)$ and $g(r)$ can be calculated from the relation
\begin{equation}
f(r) = D^{1/2}(r) F(r), \,\,\, g(r) = D^{-1/2}(r) G(r),
\label{eq:fg}
\end{equation}
where Darwin factor $D(r)$ is defined as
\begin{equation}
D(r) = \exp \left[ - 2 \int^\infty_r dr M r V_{\rm SO}(r) \right].
\end{equation}
Lower component function $G(r)$ is obtained from the relation with $F(r)$
\begin{equation}
G(r) = \frac{1}{E + M} \left[ \frac{dF}{dr} + \left( \frac{\kappa+1}{r} - T(r) \right) F \right],
\label{eq:G}
\end{equation}
where
\begin{equation}
T(r) = M r V_{\rm SO}(r).
\end{equation}
With non-relativistic nuclear potentials $V_{\rm cen}$ and $V_{\rm SO}$ given from a model,
Hartree-Fock equation is solved and $F(r)$ is obtained.
$G(r)$ is calculated from Eq.~(\ref{eq:G}), and finally the relativistic wave functions $f(r)$ and $g(r)$ are
obtained by calculating Eq.~(\ref{eq:fg}).

\subsection{KIDS model}

One purpose of the electron scattering is to get better knowledge about the structure of nuclei.
At the energy of quasi-elastic scattering, the distribution of nucleons in the interior and the gradient of density
around the surface could have effects on the cross section.
In terms of the Skyrme force, derivative terms account for the contribution of density gradient.
These terms also contribute to determining the effective mass of the nucleon,
so the structural uncertainty around the surface as well as the core could be explored by probing the dependence on the effective mass.

Role of the effective mass could be singled out when other conditions (e.g. binding energy, charge radius) are unchanged.
Independent control of the nuclear properties is accessible with the KIDS energy density functional \cite{kidsnm}.
In the KIDS framework several rules are assumed to develop a nuclear model.
At first energy per particle in homogeneous nuclear matter is expanded in powers of Fermi momentum (equally $\rho^{1/3}$) as
\begin{eqnarray}
{\cal E} (\rho,\, \delta) = {\cal T} + \sum_{i=0} (\alpha_i + \beta_i \delta^2) \rho^{1+i/3}.
\end{eqnarray}
${\cal T}$ is the kinetic energy, and $\delta = (\rho_n - \rho_p)/\rho$.
Model parameters $\alpha_i$ and $\beta_i$ are fixed or fit to symmetric and asymmetric nuclear matter properties.
In this work we fix $\alpha_0$, $\alpha_1$ and $\alpha_2$ to three saturation properties:
saturation density $\rho_0 = 0.16$ fm$^{-3}$, bindind energy per particle $E_B = 16.0$ MeV, and incompressibility $K_0 = 240$ MeV \cite{kidsnuclei2}.
Several ways have been tried to fix the $\beta_i$ values.
It has been proved that four $\beta_i$'s are good and enough to reproduce the neutron matter equation of state obtained from
microscopic calculations \cite{kidsnuclei1}, static properties of neutron rich nuclei \cite{kidsnd},
and neutron star observations \cite{kidsk0, kidssymene}.
In this work we use the values of $\beta_i$ fit to the pure neutron matter of equation of state calculated by
Akmal, Pandharipande and Ravenhall \cite{apr}.
The model thus fixed is labeled KIDS0.

\begin{table}
\begin{center}
\begin{tabular}{ccccc} \hline
 \,\,\, & \,\, KIDS0 \cite{kidsnuclei1} \,\, & \,\,  KIDS0-m*99 \,\, & \,\, KIDS-m*77 \,\, & \,\, SLy4 \cite{sly4} \, \\ \hline
$t_0$ & $-1772.04$ &  $-1772.04$ &  $-1772.04$ & $-2488.91$  \\
$y_0$ & $-127.52$ &  $-127.52$ &  $-127.52$ & $-2075.75$ \\
$t_1$ & $275.72$ & $318.92$ & $441.99$ & $486.82$ \\
$y_1$ & $0$ & $-361.17$ & $-109.03$ & $-167.37$ \\
$t_2$ & $-161.50$ & $26.82$ & $-295.06$ & $-546.39$ \\
$y_2$ & $0$ & $-215.11$ & $259.50$ & $546.39$ \\
$t_{31}$ & $12216.73$ & $12216.73$  & $12216.73$ & $13777.00$  \\
$y_{31}$ & $-11969.99$ & $-11969.99$ & $-11969.99$ & $18654.06$ \\
$t_{32}$ & $571.07$ & $-191.34$ & $-2572.65$ & - \\
$y_{32}$ & $29485.49$ & $34304.57$ & $37593.40$ & - \\
$t_{33}$ & $0$ & $0$ & $0$ & - \\
$y_{33}$ & $-22955$ & $-22955$ & $-22955$ & - \\
$W_0$ & $108.35$ & $129.96$ & $115.28$ & 123.0 \\ \hline
$\mu_S$ & 0.99 & 0.90 & 0.70 & 0.70 \\
$\mu_V$ & 0.81 & 0.90 & 0.70 & 0.80  \\
\hline
\end{tabular}
\end{center}
\caption{Parameters in the single particle potential Eqs. (\ref{eq:vcen}, \ref{eq:vso}).
We use the simplified notation $y_i = t_i x_i$. Units of the parameters are
MeV$\cdot$fm$^{3}$ for $t_0$, $y_0$,
MeV$\cdot$fm$^{4}$ for $t_{31}$, $y_{31}$,
MeV$\cdot$fm$^{5}$ for $t_1$, $y_1$, $t_2$, $y_2$, $t_{32}$, $y_{32}$, $W_0$,
and 
MeV$\cdot$fm$^{6}$ for $t_{33}$, $y_{33}$.
Isoscalar and isovector effective masses in each model are shown in the last two rows.}
\label{tab:para}
\end{table}

When nuclei are described in the KIDS framework, energy density functional is transformed to the form of Skyrme force \cite{kidsnpsm2017}.
Terms accounting for the density gradient and spin-orbit interactions are added 
In the notation of Skyrme force, they correspond to parameters $t_1$, $t_2$, $x_1$, $x_2$ and $W_0$.
We assume $t_1=t_2$, $x_1=x_2=0$, and $t_1$ and $W_0$ are fit to 
the data of binding energy and charge radius of
$^{40}$Ca, $^{48}$Ca and $^{208}$Pb.
In the KIDS0 model, $x_1$ and $x_2$ are assumed to be 0, so the effective masses are obtained as results of determining $t_1$.
By adjusting $x_1$ and $x_2$, one can produce specific values of isoscalar and isovector effective masses 
$\mu_S$ and $\mu_V$ defined by
\begin{eqnarray}
\mu_S &=& \frac{m^*_S}{M} = \left[1 + \frac{M}{8 \hbar^2} \rho (3 t_1 + 5 t_2 + 4 y_2) \right]^{-1}, \nonumber \\
\mu_V &=& \frac{m^*_V}{M} = \left[ 1 + \frac{M}{4 \hbar^2} \rho (2 t_1 + 2 t_2 + y_1 + y_2) \right]^{-1}, 
\end{eqnarray}
where $y_i = t_i x_i$.
In this work we consider two cases $(\mu_S,\, \mu_V) = (0.7, 0.7)$ and $(0.9, 0.9)$.
Each model is labeled KIDS0-m*77 and KIDS0-m*99, respectively.

Central and spin-orbit potentials entering Eq. (\ref{eq:Fr}) are obtained from KIDS functional as
\begin{eqnarray}
V_{\text{cen}} &=& U_q + U_{\text{coul}}, \nonumber \\
U_q &=& \left(t_0 + \frac{1}{2} y_0\right)\rho - \left(\frac{1}{2} t_0 + y_0\right)\rho_q \nonumber \\
&+& \frac{1}{8}[(2t_1+y_1)+(2t_2+y_2)]\tau - \frac{1}{8}[(t_1+2y_1)+(t_2+2y_2)]\tau_q \nonumber \nonumber \\
&-& \frac{1}{16}[(6t_1+y_1)-(2t_2+y_2)]\nabla^2 \rho +\frac{1}{16}[(3t_1+6y_1)+(t_2+2y_2)]\nabla^2 \rho_q \nonumber \\
&+&  \frac{1}{12} \sum_{k=1}^3 \rho^{k/3}\left[\left(2+\frac{k}{3}\right)\left(t_{3k} + \frac{1}{2}y_{3k}\right)\rho - (t_{3k}+2y_{3k})\rho_q 
-\frac{k}{3} \left(\frac{1}{2} t_{3k} + y_{3k}\right) \frac{\rho^2_{p}+\rho^2_{n}}{\rho^2}\right] \nonumber \\
&-& \frac{W_0}{2}\left(\frac{\partial}{\partial r} + \frac{2}{r} \right) (J + J_q), 
\label{eq:vcen} \\
V_{\text{SO}} &=& \frac{1}{2}W_0 \left(\frac{\partial}{\partial r} + \frac{2}{r} \right) (\rho + \rho_q) 
+ \frac{1}{8}(t_1-t_2)J_q-\frac{1}{8}(y_1+y_2)J, 
\label{eq:vso}
\end{eqnarray}
where $\tau$ is the non-relativistic kinetic energy, $J_q$ denotes the spin current of the neutron and the proton, and $J = J_n + J_p$.
Parameters in the potential of each model are summarized in Tab. \ref{tab:para}.
We consider the SLy4 model for a comparison with a standard Skyrme force model. 
With these potentials, the transformed relativistic wave functions are generated and applied into the exclusive
$(e, e' p)$ reaction in quasi-elastic region.

\section{Result and Discussion}

\subsection{Density distribution}

\begin{figure}[t]
\begin{center}
\includegraphics[width=1.0\textwidth]{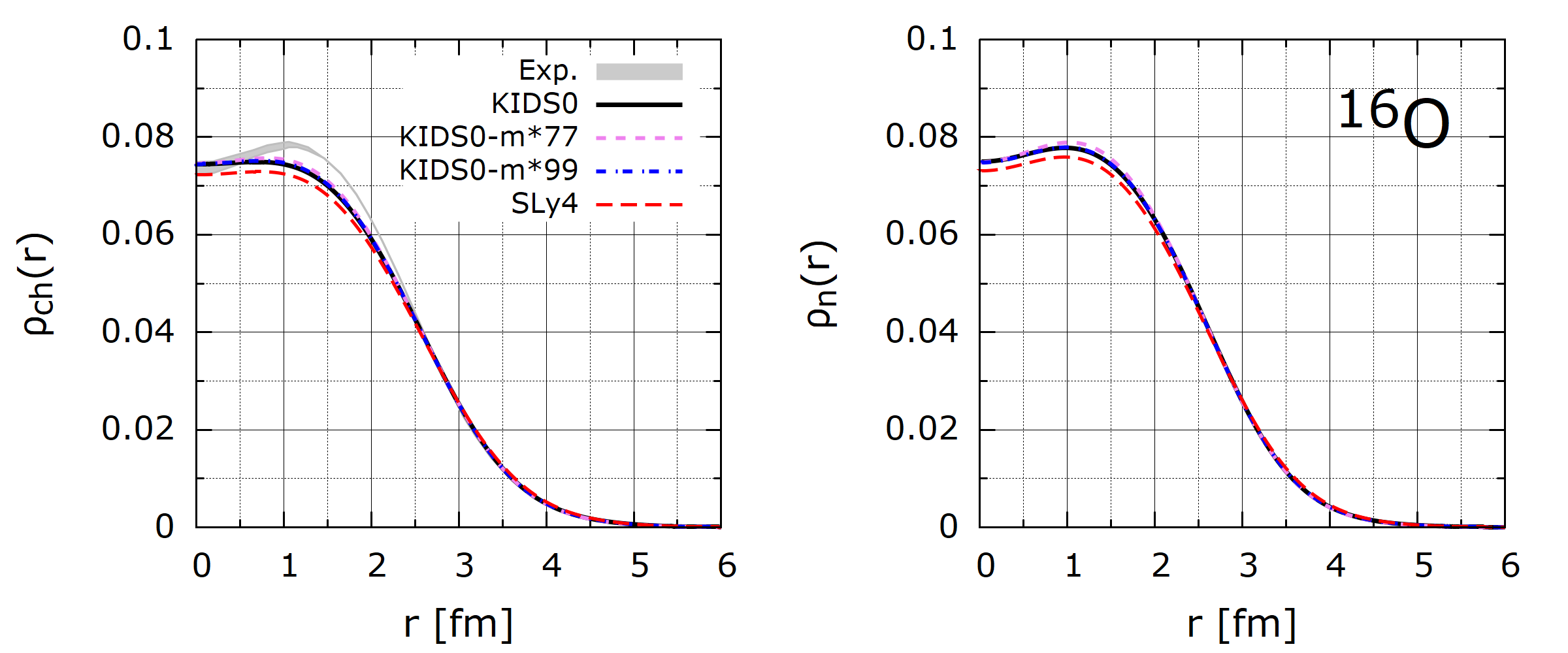}
\end{center}
\caption{Density profile of charge (left) and neutron (right) for $^{16}$O.}
\label{fig:density-16}
\end{figure}

\begin{figure}[t]
\begin{center}
\includegraphics[width=1.0\textwidth]{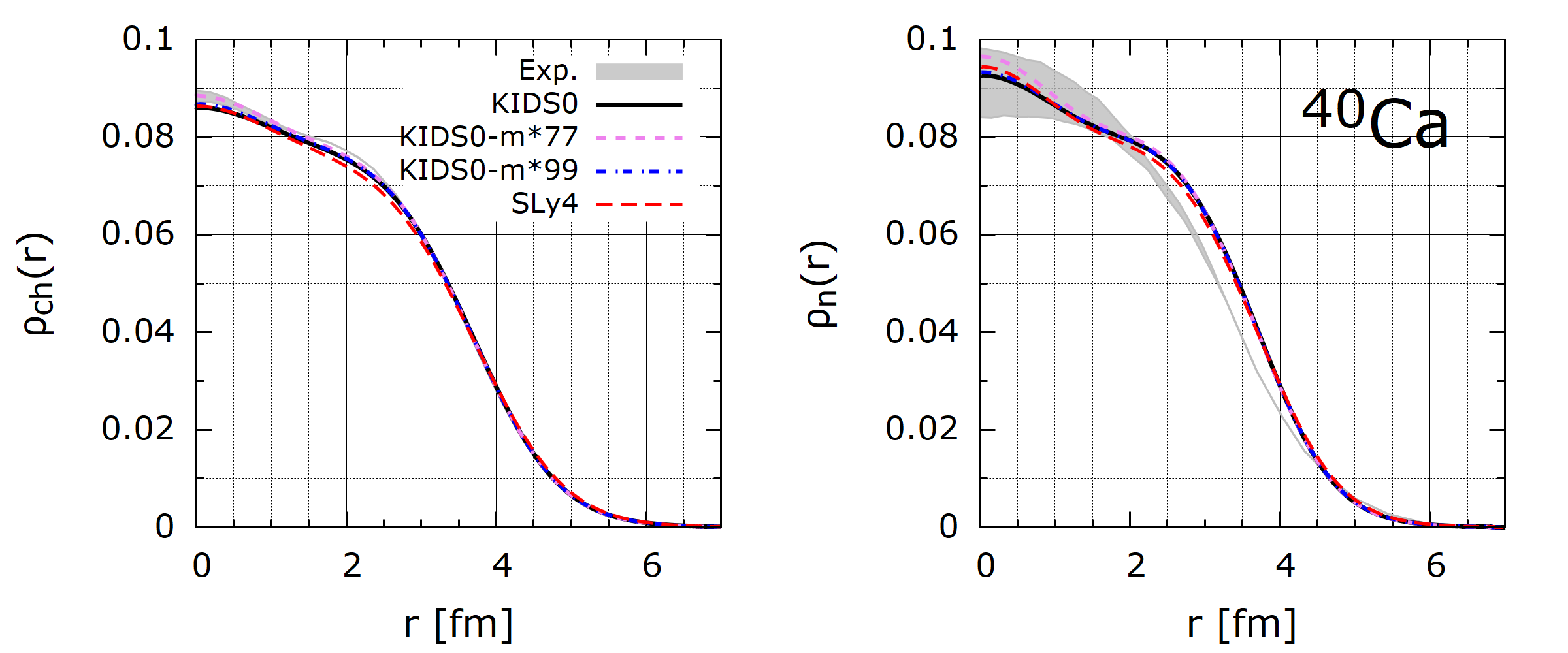}
\end{center}
\caption{Density profile of charge (left) and neutron (right) for $^{40}$Ca.}
\label{fig:density-40}
\end{figure}

\begin{figure}[t]
\begin{center}
\includegraphics[width=1.0\textwidth]{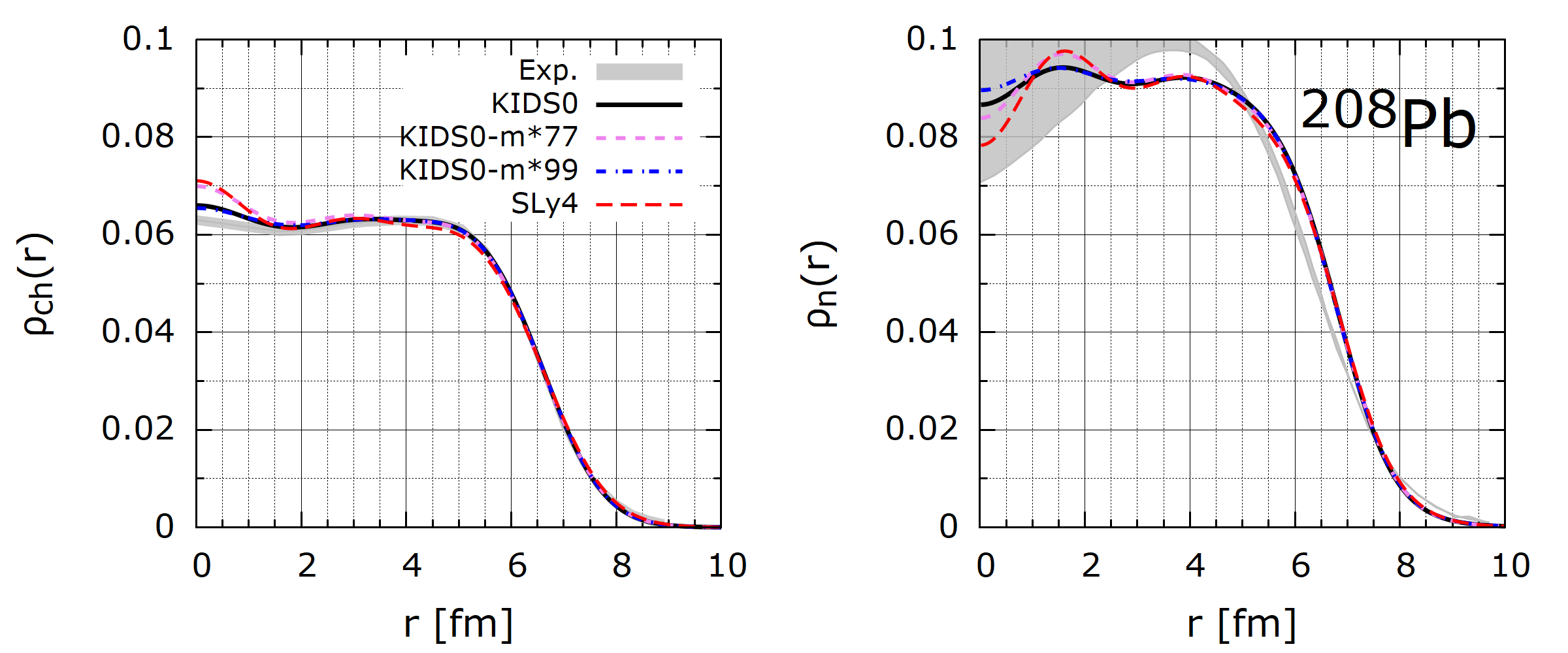}
\end{center}
\caption{Density profile of charge (left) and neutron (right) for $^{208}$Pb.}
\label{fig:density-208}
\end{figure}

Charge and neutron distributions in $^{16}$O, $^{40}$Ca and $^{208}$Pb are
displayed in Figs. \ref{fig:density-16}, \ref{fig:density-40}, and \ref{fig:density-208}, respectively.
Measured values are denoted with gray bands.

In the result of $^{16}$O, KIDS models show weak dependence on the effective mass, 
and the results of models agree to each other in both charge and neutron distributions.
SLy4 model agrees well with KIDS model at $r > 2$ fm, but in the interior region ($r<2$ fm)
densities are slightly suppressed compared to the KIDS model.
All the models reproduce the data of charge distribution \cite{o16density} well over $r>$ 2.5 fm.
Visual discrepancy is found in $0.5 < r < 2.5$ fm, but the difference from experiment is less than 10\%.
Data for the neutron are not available so only the theory results are presented.

For $^{40}$Ca, model dependence is weak again, and the four models predict similar distributions of both protons and neutrons.
In the comparison with experiment \cite{sick1981, ray1979}, distribution of the proton is reproduced well by the theory.
In case of the neutron, theoretical results are within the errors of experiment in the core region ($r<2$ fm),
but on the surface ($2<r<4$ fm) where the density drops rapidly, theory exceeds experiment.

We have seen that  the distribution of the proton and the neutron are insensitive to the effective mass in the light nuclei $^{16}$O and $^{40}$Ca.
In the result of $^{208}$Pb, on the other hand, the effect of effective mass appears clear.
For the proton, dependence on the effective mass is negligible at $r>1$ fm.
In this region model dependence is weak and theory agrees well with experiment \cite{pb208density}.
At $r<1$ fm, theories are divided into two groups: one with KIDS0, KIDS0-m*99 (Group1),
and the other with KIDS0-m*77, SLy4 (Group2).
Group2 shows charge density large than Group1 as $r \rightarrow 0$. 
Models in each group have similar isoscalar effective mass, $\mu_S \sim 1.0$ for Group1 and $\mu_S=0.7$ for Group2.
Therefore different behavior of charge density at $r < 1$ fm could be originated from the value of isoscalar effective mass.

Neutron distribution shows pattern of agreement and disagreement to experiment \cite{saito2007} similar to the neutron distribution of $^{40}$Ca.
Dependence on the effective mass is divided into three categories: no dependence at $r > 2.5$ fm,
models are classified into two sets Group1 and Group2 at $1 < r < 2.5$ fm, and 
four models behave independently at $r<1$ fm.
In the inner core $r<2$ fm, theory results are within the range of experimental uncertainty.
In the outer core region $2 < r < 5$ fm, all the models obtain neutron densities less than experiment.
The discrepancy does not exceed 10\%.
In the surface region $5<r<8$ fm, theory overwhelms experiment slightly, and it is reversed in the tail $r>8$ fm.
There seems to be a pattern in the discrepancy of the neutron distribution.
On the other hand, distribution of the proton which is relevant to the scattering with electrons agrees well with experiment.

\subsection{Single particel levels}

\begin{table}[t]
\begin{center}
\begin{tabular}{c|cc|cc|cc}\hline
 & \multicolumn{2}{c|}{$^{16}$O} & \multicolumn{2}{c|}{$^{40}$Ca} & \multicolumn{2}{c}{$^{208}$Pb} \\ 
 &  1p$_{1/2}$ & 1p$_{3/2}$ & 2s$_{1/2}$ & 1d$_{3/2}$ & 3s$_{1/2}$ & 2d$_{3/2}$ \\ \hline
Exp. \cite{volya2007} & $-12.13$ & $-18.40$ & $-10.92$ & $-8.33$ & $-8.01$ & $-8.36$ \\ 
KIDS0 & $-9.67$ & $-14.76$ & $-8.90$ & $-7.26$ & $-8.21$ & $-8.96$ \\ 
KIDS0-m*99 & $-9.42$ & $-15.42$ & $-9.10$ & $-6.95$ & $-8.47$ & $-9.24$ \\ 
KIDS0-m*77 & $-10.99$ & $-16.53$ & $-9.81$ & $-8.49$ & $-8.64$ & $-9.64$ \\ 
SLy4 & $-10.53$ & $-15.99$ & $-9.82$ & $-8.08$ & $-8.79$ & $-9.57$ \\ \hline
\end{tabular}
\end{center}
\caption{Single particle levels in MeV.}
\label{tab:spl}
\end{table}

Electron-nucleus scattering data provide cross sections from protons at specific states.
Table \ref{tab:spl} collects the single particle levels of the proton for which
cross sections calculated from theory will be compared with experiment.
For light nuclei $^{16}$O and $^{40}$Ca, models with smaller isoscalar effective mass (Group2)
obtain results closer to experiment than the models in Group1.
For $^{16}$O, Group2 models differ from experiment by 9.4--13.2\%, and Group1 models by 16.6--18.5\%.
For $^{40}$Ca, Group1 models give deviations from experiment 16.6--18.5\% and 12.8--16.6\% in the 2s$_{1/2}$ and 1d$_{3/2}$ states, respectively.
With the models in Group2, we have 10.1--10.2\% and 2--3\% for 2s$_{1/2}$ and 1d$_{3/2}$ states, respectively.

For $^{208}$Pb Group1 models show better agreement to data.
In the 3s$_{1/2}$ state Group1 models give differences 2.5--5.7\% and Group2 models 7.9--9.7\%.
In the 2d$_{3/2}$ state, Group1 and Group2 give differences 7.2--10.5\% and 14.5--15.3\%, respectively.
A similar pattern is reported in the UNEDF model \cite{unedf0}.
Isoscalar effective mass is $0.9 M$ in the UNEDF model,
and the model shows agreement better for heavy nuclei than light ones.

As far as single particle levels are concerned, light nuclei favor small isoscalar effective mass, 
but heavy nuclei support effective masses close to 1.

\subsection{Cross section}

\begin{figure}[t]
\begin{center}
\includegraphics[width=1.0\textwidth]{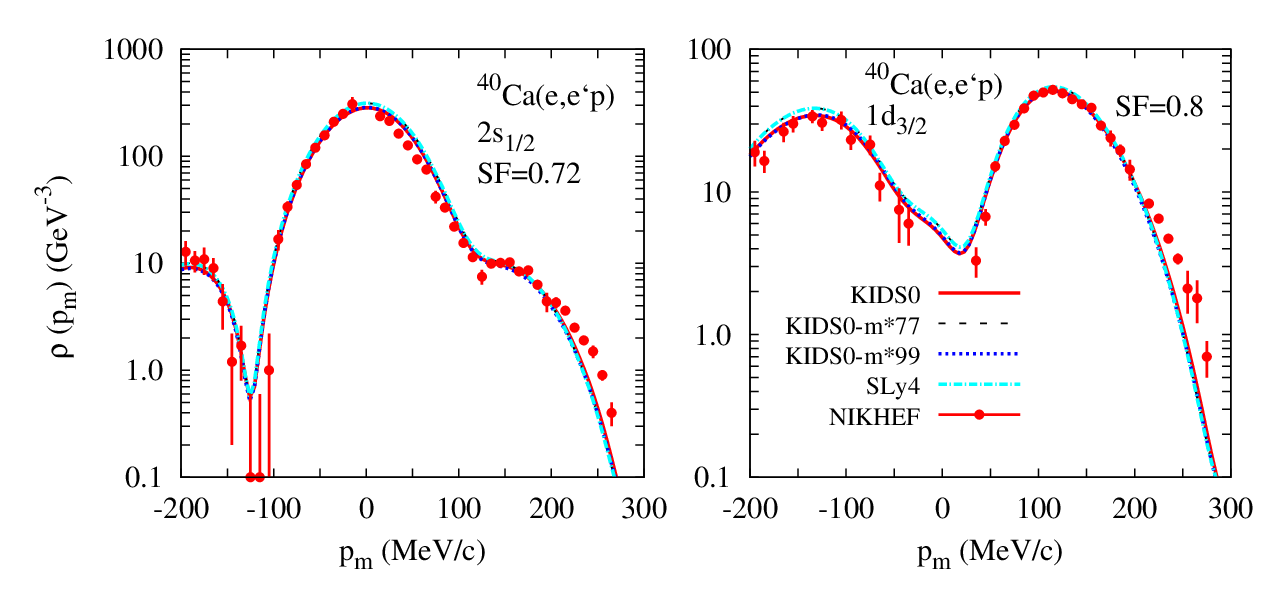}
\end{center}
\caption{$^{40}$Ca cross section with parallel kinematics. 
The experimental data taken from NIKHEF \cite{kramer}.}
\label{fig:ca-par}
\end{figure}

In the present work, we calculate the reduced cross section $\rho(p_m)$ at one particular shell,
which is related to the probability that a bound nucleon at a given orbit with the missing
momentum $\mathbf{p}_m$ can be knocked out of the nucleus with asymptotic momentum $\mathbf{p}$.
The reduced cross section as a function of $p_m$ is commonly defined by
\begin{eqnarray}
\rho(p_m) = {\frac {1} {pE \sigma_{ep}}} {\frac {d^3 \sigma} {dE_f d\Omega_f d\Omega_p}},
\label{eq:reduced_cr}
\end{eqnarray}
where the missing momentum is determined by the kinematics $\mathbf{p}_m = \mathbf{p}-\mathbf{q}$
with $\mathbf{q}$ the momentum of virtual photon mediating EM interactions between electron and proton.
The off-shell electron-proton cross section, $\sigma_{ep}$ is not uniquely defined.
We use the form CC1 $\sigma_{ep}$ given by Ref. \cite{npa1983}.
In all the calculations SFs are calibrated with the KIDS0 model.

Figure \ref{fig:ca-par} shows the cross section of $^{40}$Ca in the parallel kinematics where $\mathbf{p} \parallel \mathbf{q}$.
The incident electron energy is $E_i=412$ MeV and the energy transfer to proton is $\omega=100$ MeV.
Theory results are represented with lines, and data from NIKHEF \cite{kramer} are noted with red circles.
In the 2s$_{1/2}$ state, theory results are similar to each other.
Spectroscopic factors are adjusted to reproduce the cross section data around peak at $p_m = 0$ MeV/$c$ with KIDS0 model.
Theory agrees well with data not only at $p\simeq 0$ MeV/$c$, but over $p_m < 200$ MeV/$c$.
In the 1d$_{3/2}$ state, experiment at peaks around $p_m \simeq -150$ MeV/$c$ and 100 MeV/$c$ are reproduced with good accracy,
and agreement is extended over the range $p_m < 200$ MeV/$c$.
Model dependence is weak, but a small gap between Group1 and Group2 is seen at $p_m <0$.
Group2 models predict the cross sections slightly larger than the Group1 models.

\begin{figure}[t]
\begin{center}
\includegraphics[width=1.0\textwidth]{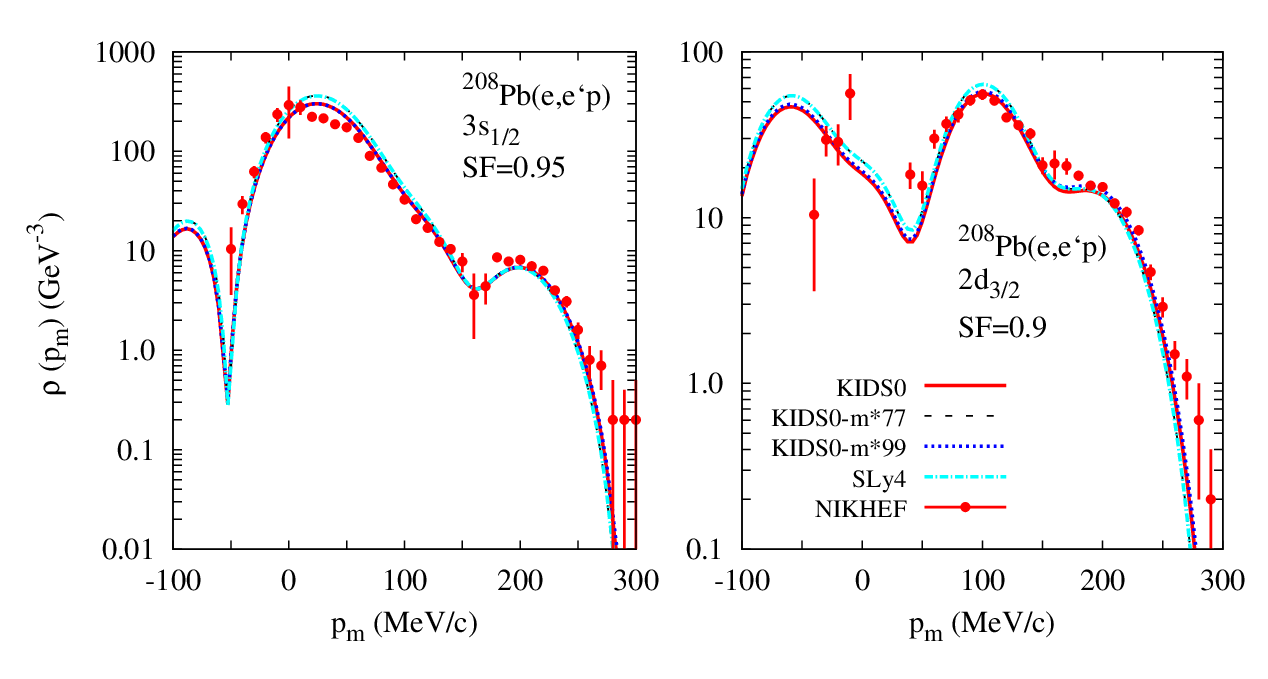}
\end{center}
\caption{$^{208}$Pb cross section with parallel kinematics.
The experimental data were measured from NIKHEF \cite{bobeldijk}.}
\label{fig:pb-par}
\end{figure}

Figure \ref{fig:pb-par} presents the result of $^{208}$Pb in the parallel kinematics.
The incident electron energy is $E_i=412$ MeV and the kinetic energy of knocked-out proton is $T_p=100$ MeV.
Predictions are divided into two groups around the peaks in both 3s$_{1/2}$ and 2d$_{3/2}$ states.
Since the SFs are calibrated with KIDS0 model, models in Group1 are in good agreement with the NIKHEF data \cite{bobeldijk}.
Models in Group2 that have $\mu_S=0.7$ give cross sections larger than the Group1 models by 15--30\%.
Isovector effective mass does not follow the grouping of isoscalar effective mass:
KIDS0 and SLy4 models have $\mu_V=0.8$, KIDS0-m*77 has $\mu_V=0.7$, and $\mu_V=0.9$ for the KIDS0-m*99 model.
Therefore it is likely that the model dependence of the cross section is dominated by the isoscalar effective mass,
and the effect of the isovector effective mass is, if ever, quite limited.
Result of $\mu_S=0.9$ (KIDS0-m*99) can hardly be differentiated from that of $\mu_S=1.0$ (KIDS0) in the 3s$_{1/2}$ state, 
but in the 2d$_{3/2}$ state, result of the former is slightly enhanced over the latter.
This difference is, though small, in accordance with the correpondence of small effective mass to large cross section.

\begin{figure}[t]
\begin{center}
\includegraphics[width=1.0\textwidth]{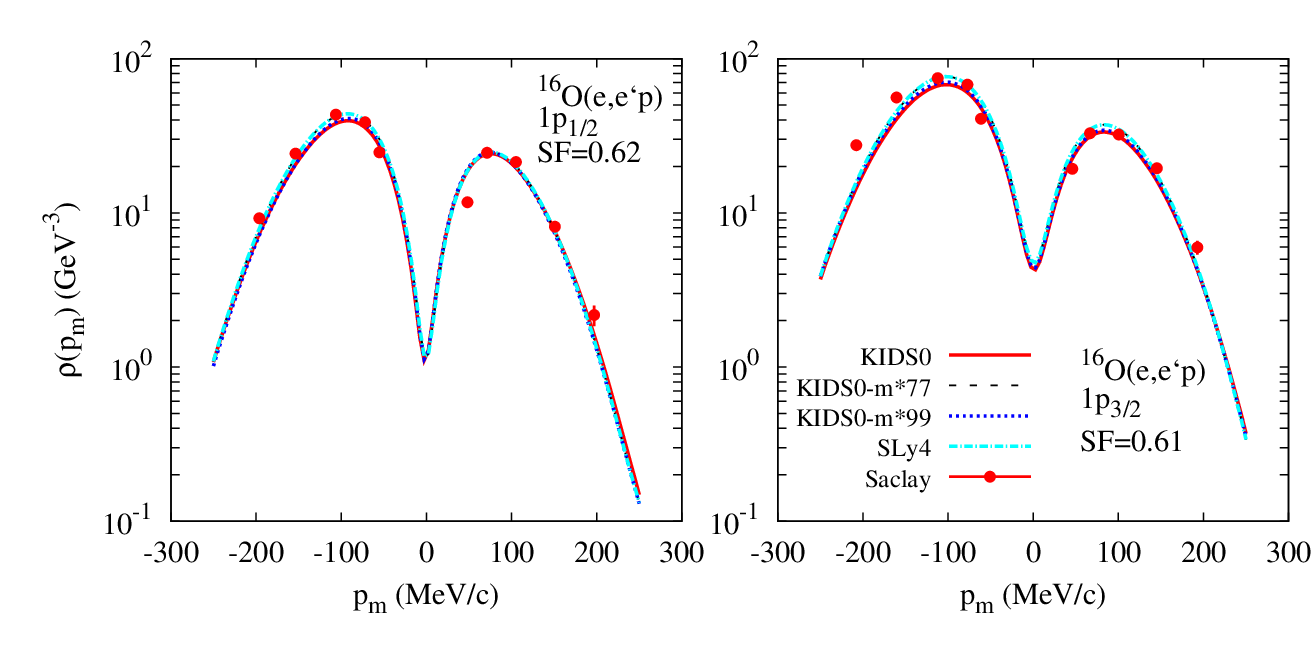}
\includegraphics[width=1.0\textwidth]{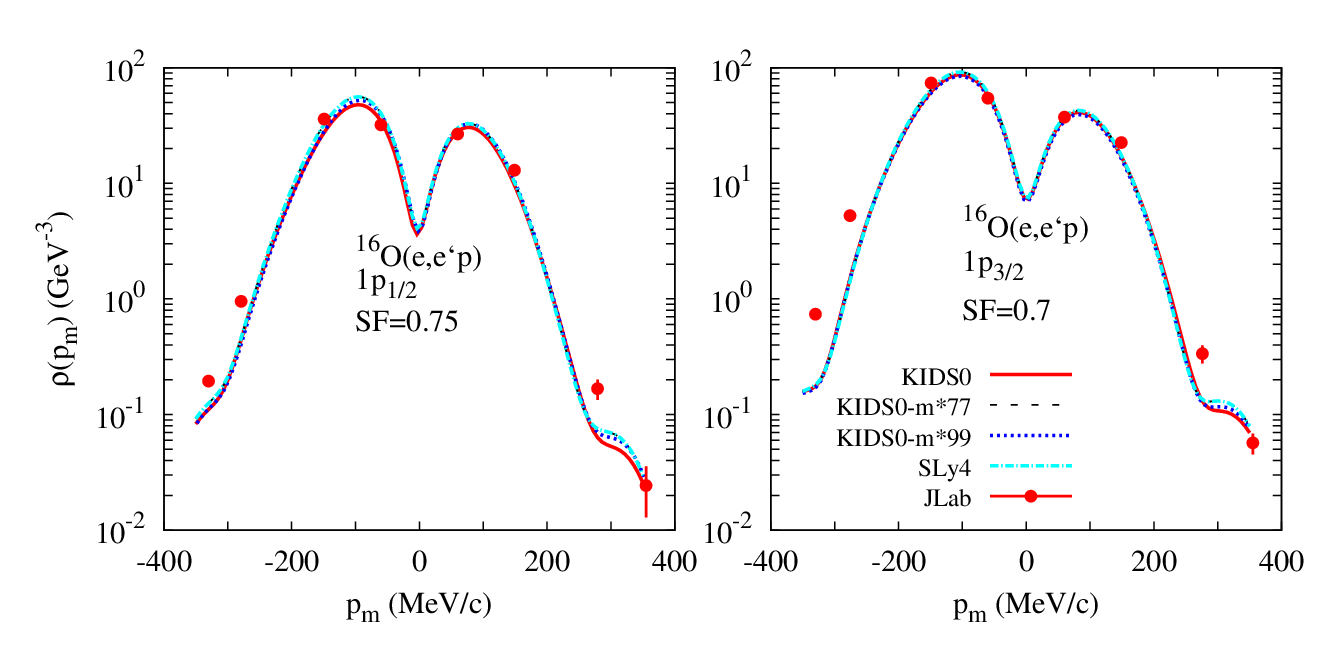}
\end{center}
\caption{$^{16}$O cross section with perpendicular kinematics.
The experimental data in the upper panels were measured from Saclay \cite{chinitz} and in the lower panels from JLab \cite{gao}.}
\label{fig:o-per}
\end{figure}

Figure \ref{fig:o-per} depicts the result of $^{16}$O in the perpendicular kinematics.
In the perpendicular kinematics $|\mathbf{p}| = |\mathbf{q}|$, so $\mathbf{p}_m$ is almost perpendicular to $\mathbf{q}$.
Upper panels compare the result with data from Saclay \cite{chinitz}, and the lower panels with data from JLab \cite{gao}.
Energies of the incident electrons are 580 MeV in Saclay and 2441 MeV in JLab.
The kinetic energies of the knocked-out protons are $T_p=159$ MeV and $T_p=427$ MeV, respectively.
As a result, even though the ranges of $p_m$ are similar,
both theoretical results and experimental data are different in the upper and lower panels.
For the same reason, SFs are different depending on the incident electron energies.
Results obtained from the models are not sensitive to the effective mass,
so they agree well with experiment at both energies.
Looking into the details around peaks, cross sections are increasing in the order of KIDS0, KIDS0-m*99
and KIDS0-m*77. Results of SLy4 are indistinguishable from those of KIDS0-m*77.
Again the isoscalar effective masses are decreasing in the order of increasing cross sections.
This behavior is consistent with what has been observed in the result of $^{208}$Pb.
The result supports that the correlation between the isoscalar effective mass and the scattering cross section 
is not limited to specific nuclei, but valid over a wide range of mass number. 

\begin{figure}[t]
\begin{center}
\includegraphics[width=1.0\textwidth]{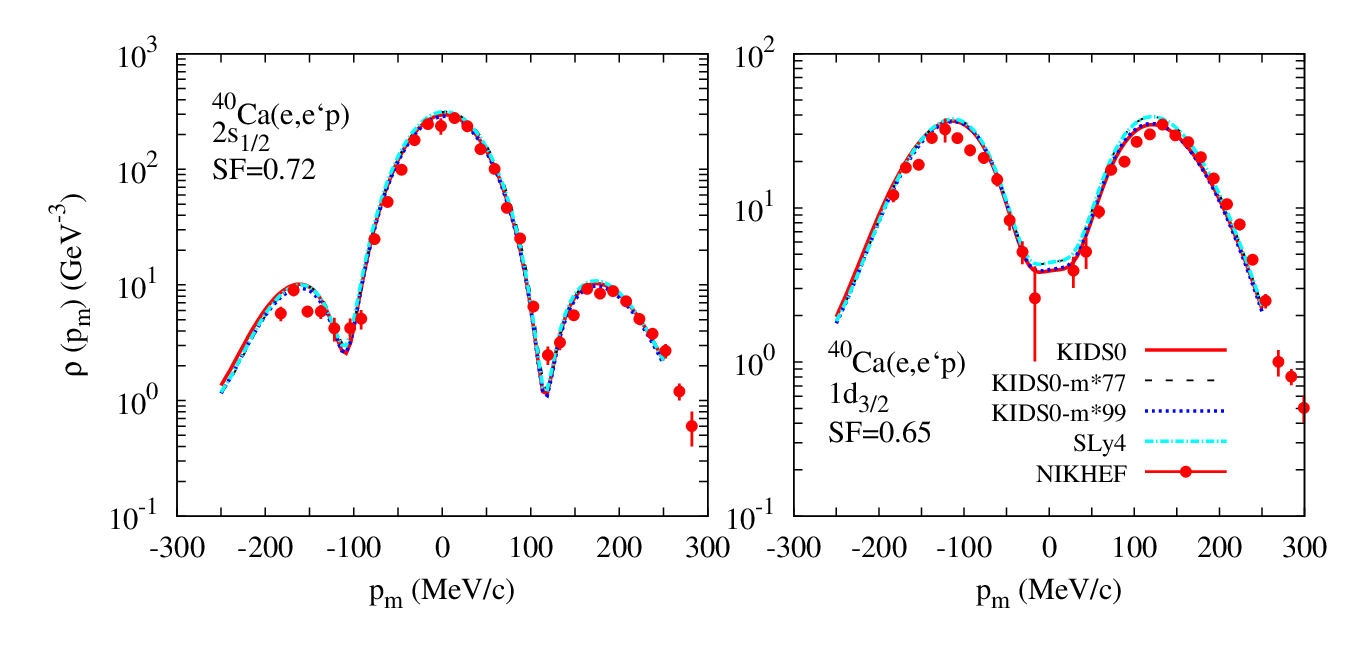}
\end{center}
\caption{$^{40}$Ca cross section with perpendicular kinematics.
The experimental data were measured from NIKHEF \cite{kramer}.}
\label{fig:ca-per}
\end{figure}

Figure \ref{fig:ca-per} reports the cross section of $^{40}$Ca in the perpendicular kinematics
with the incident electron energy and the energy transfer to proton the same with those in Fig. \ref{fig:ca-par}.
Similar to the parallel kinematics, theory reproduces the data with accracy,
and the dependence on the effective mass is marginal.
Spectroscopic factors are the same with that of the parallel kinematics in the 2s$_{1/2}$ state,
but we have a reduced value in the 1d$_{3/2}$ state.
In the result of $^{16}$O, we saw that the kinematic conditions have effect to the value of SF.
Different SF values in the 1d$_{3/2}$ state could be attributed to the kinematic conditions.

\begin{table}[tbp]
\begin{center}
\begin{tabular}{ccccccc}\hline
 & &\,\, This work \,\, & \,\, Ref. \cite{jin1992}\,\,  &\,\,  Ref. \cite{udias1993}\,\,  
&\,\,  Ref. \cite{volya2007}\,\,  &\,\,  Ref. \cite{gnez2014} \\ \hline
$^{40}$Ca & 2s$_{1/2}$ & 0.72            & 0.75 & 0.44 -- 0.51 & 0.87 & 0.825 -- 0.931\\ 
                & 1d$_{3/2}$ & 0.65 -- 0.80 & 0.80 & 0.60 -- 0.76 & 0.93 & 0.848 -- 0.966 \\ \hline
$^{208}$Pb & 3s$_{1/2}$ & 0.95           & 0.71 & 0.65 -- 0.70 & 0.85 & 0.787 -- 0.929 \\
                 & 2d$_{3/2}$ & 0.90           & -    & 0.66 -- 0.73 & 0.90 & 0.783 -- 0.937 \\ \hline
\end{tabular}
\end{center}
\label{tab:sf}
\caption{Spectroscopic factors calculated from theory \cite{jin1992, udias1993, gnez2014} and extracted from experiment \cite{volya2007}.}
\end{table}

Spectroscopic factors provide understanding of the structural details of nuclei, 
and allow the estimation of the contribution of many-body correlations that could be missed in the mean field approximation.
Therefore SFs is a measure to figure out the validity, accuracy and limit of shell description of a model. 
Table III collects SFs from literature, and compares them with the result of this work.
Calculations based on relativistic formalism \cite{jin1992, udias1993} obtain results smaller than the values
evaluated from experiment \cite{volya2007} and a phonon-coupling calculation \cite{gnez2014}.
Results of the latter two \cite{volya2007, gnez2014} agree to each other.
Result of $^{208}$Pb in this work is consistent with Refs. \cite{volya2007, gnez2014},
but those of $^{40}$Ca agree with relativistic calculations \cite{jin1992, udias1993}.
Consequently our work predicts SFs of $^{40}$Ca smaller than those of $^{208}$Pb.
It could be interpreted that the mean field approximation works better in heavy nuclei.
Result of $^{16}$O is consistent with this tendency because the SFs are 0.62 -- 0.75 in 1p$_{1/2}$ and 0.61 -- 0.70 in 1p$_{3/2}$.

\subsection{Response function}

\begin{figure}[t]
\includegraphics[width=12cm]{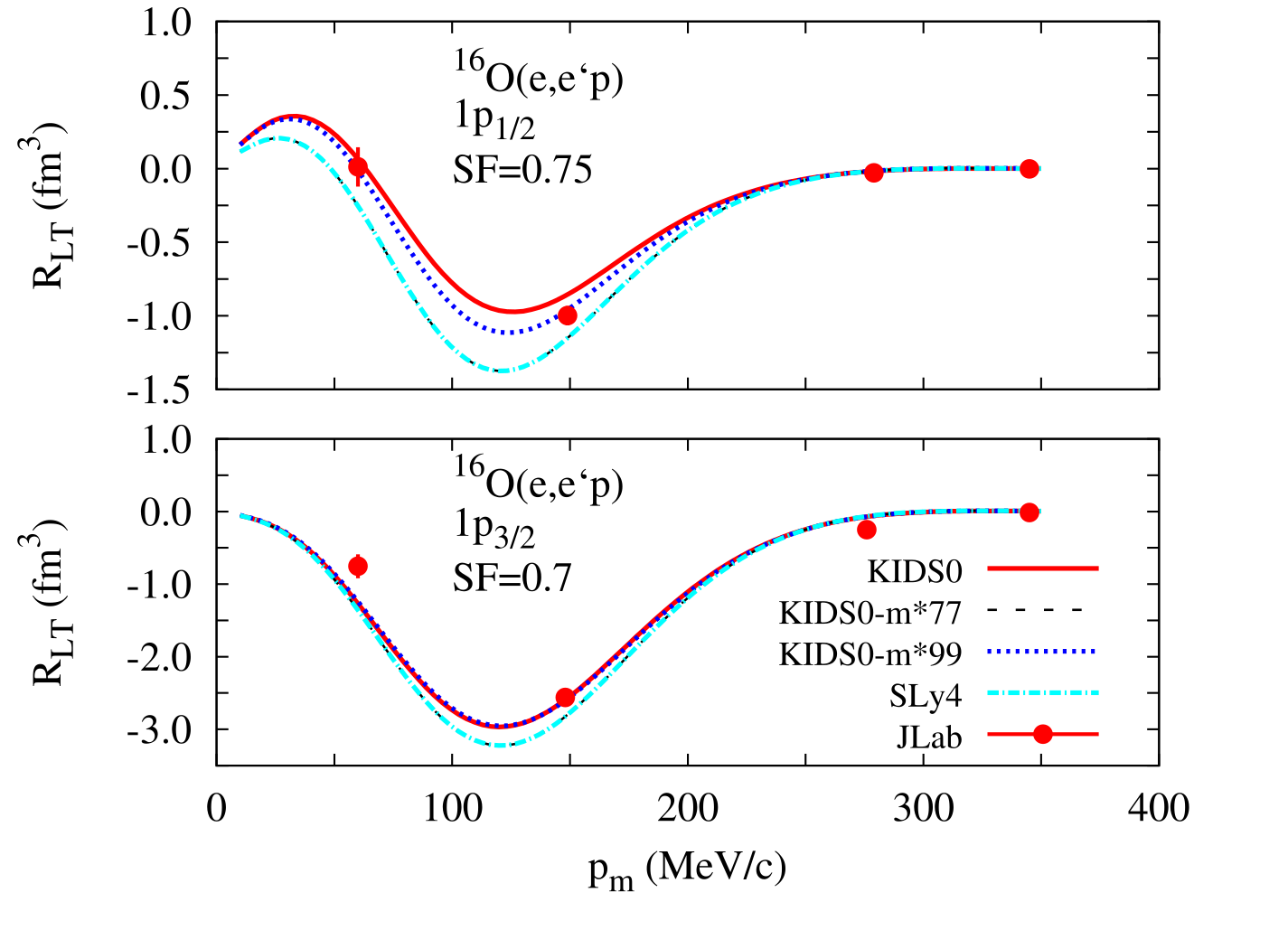}
\caption{The fourth response functions from $^{16}$O with the kinematics same as the JLab.}
\label{fig8}
\end{figure}

In the laboratory frame, the quasi-elastic cross section for the $(e, e' p)$ reaction is simply written as
\begin{eqnarray}
{\frac {d^3 \sigma} {dE_f \,d\Omega_f \, d\Omega_p} } &=& K  \Big[ v_{L} R_{L}  + v_{T} R_{T} + \cos2\phi_p \,v_{TT} R_{TT} \nonumber \\
&&\qquad\qquad\qquad + \cos\phi_{p}\, v_{LT} R_{LT} + h \sin\phi_p\, v_{LT'} R_{LT'} \Big]~,
\label{cs}
\end{eqnarray}
where the kinematic factor $K$ is given by $K={\frac {p \,E} {(2\pi)^3}}\, \sigma_{M}$ with the Mott cross section 
$\sigma_M = {\frac {\alpha^2} {4E^2_i}}\, {\frac {\cos^2({\theta_e}/2)} {\sin^4({\theta_e}/2)}}$.
In Eq.~(\ref{cs}), $R_{L}$, $R_T$, $R_{TT}$, and $R_{LT}$ are referred to as the longitudinal, transverse,  longitudinal-transverse, 
and transverse-transverse interferences response functions, respectively.
The fifth one, $R_{LT'}$, indicates the polarized longitudinal-transverse interference, which is directly proportional to the electron beam asymmetry.
The four-momenta of the incoming and outgoing electrons are labeled $p_i^{\mu}=(E_i, {\bf p}_i)$ and $p_f^{\mu}=(E_f, {\bf p}_f)$.
In the parallel kinematics, three interference terms disappear, but they can be extracted in the perpendicular kinematics.
The detailed discussions are in Ref. \cite{kimepja01}.
The electron kinematic factors in Eq.~(\ref{cs}) are given in terms of the four-momentum transfer, $q=(\omega,\mathbf{q})$, 
and the electron scattering angle $\theta_e$:
\begin{eqnarray}
	v_{L} = \frac{q^4}{\mathbf{q}^4}~, \qquad v_{T} = \tan^2\frac{\theta_e}{2} - \frac{q^2}{2\mathbf{q}^2}~, \qquad
	v_{TT} = - \frac{q^2}{2\mathbf{q}^2}~,\nonumber \\
	v_{LT} = -\frac{q^2}{\mathbf{q}^2}\,\left( \tan^2\frac{\theta^{}_e}{2} - \frac{q^2}{2\mathbf{q}^2}  \right)~,\qquad
	v_{LT'} = -\frac{q^2}{\mathbf{q}^2}\, \tan^2\frac{\theta_e}{2}.
\end{eqnarray}

In Eq.~(\ref{cs}), the fourth structure function could be obtained by subtracting the cross sections at
azimuthal angles of the outgoing proton $\phi_p = 0$ and $\phi_p=\pi$ and keeping the other electron 
and outgoing proton kinematics variables fixed. The fourth structure function is a function of the missing momentum given by
\begin{equation}
R_{LT} = {\frac {\sigma^R - \sigma^L} {2Kv_{LT}}} ,
\end{equation}
where $L$ (left) and $R$ (right) indicate the left side at $\phi_p=0$ and the right side at $\phi_p=\pi$ of the cross section in Eq. (\ref{cs}), respectively.

\begin{figure}[t]
\includegraphics[width=12cm]{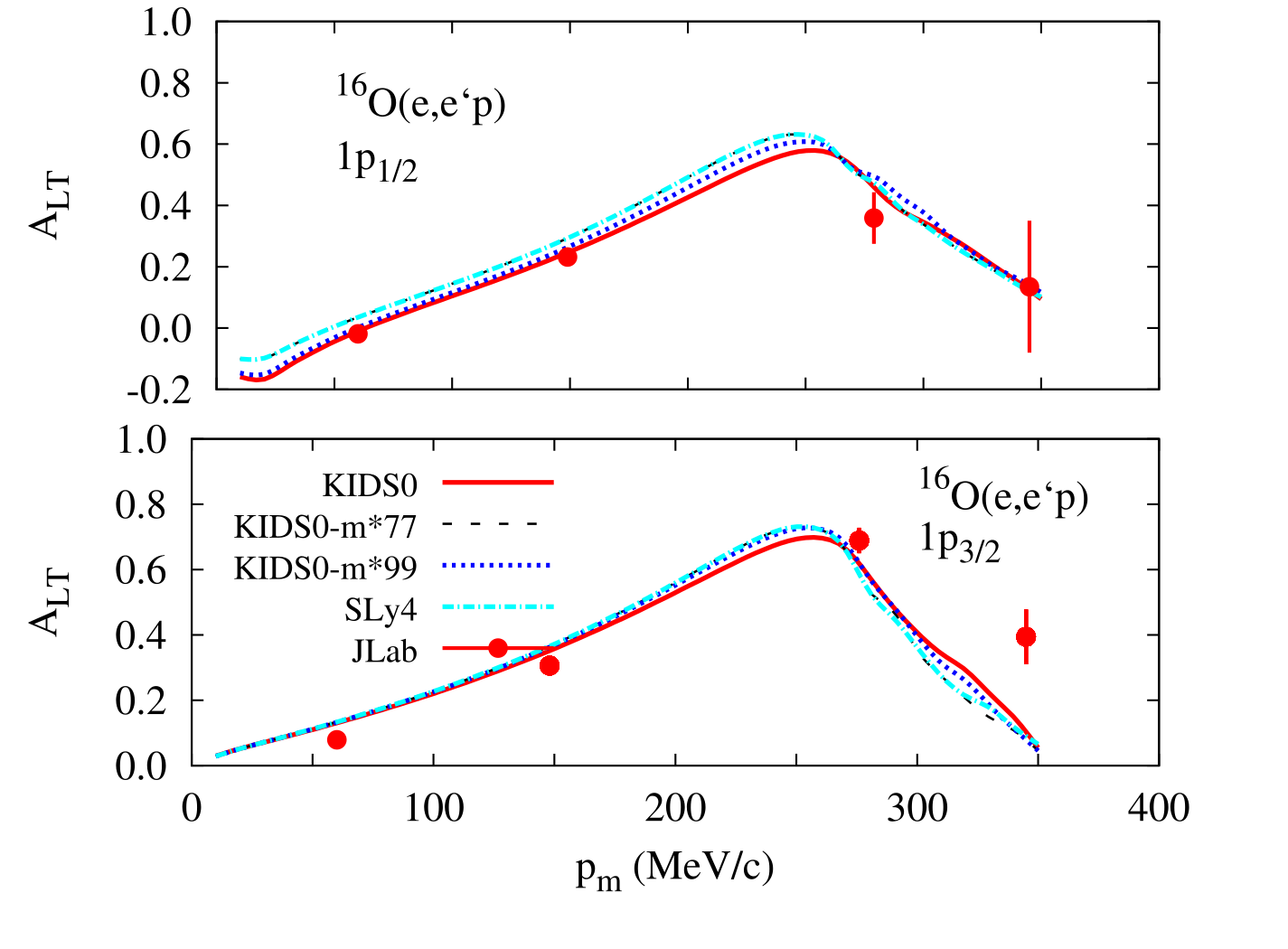}
\caption{The left-right asymmetry from $^{16}$O with the kinematics same as the JLab.}
\label{fig9}
\end{figure}

Figure \ref{fig8} shows the fourth response functions from $^{16}$O with the same kinematics as the JLab experiment. 
The explanations of the curves are the same as the previous results. 
The difference between the effective masses is at most about 45 \%  and 10 \% around the position of peak at 1p$_{1/2}$ and 1p$_{3/2}$ orbits, respectively. 
The sensitivity of the effective masses is clearly exhibited on the fourth response function.
We learn that the values of the SF for
the fourth response function and for the reduced cross section may be different for a good
agreement with the data.

Finally, we also calculate another left-right asymmetry, $A_{LT}$, defined as
\begin{equation}
A_{LT} = {\frac {\sigma^R - \sigma^L} {\sigma^R + \sigma^L}}.
\end{equation}
The kinematics in Fig. \ref{fig9} are the same as in Fig. \ref{fig8}. 
Since this asymmetry does not require any SF, it is possible to compare the theoretical result with experimental data directly. 
The role of the effective masses is not large compared to the fourth response function but our theoretical results describe 
the experimental data relatively well except one point at $p_m \simeq 350$ MeV/$c$ in the 1p$_{3/2}$ orbit.

\section{Summary}

In the present work, we have explored the ground state properties of spherical nuclei with KIDS model
by considering the quasi-elastic scattering of electrons from $^{16}$O, $^{40}$Ca, and $^{208}$Pb.
Wave functions of the nucleon in nuclei were obtained by solving non-relativistic Hartree-Fock equations.
They were transformed to a relativistic form, and the cross sections were calculated in the relativistic formalism.
The effects of the isoscalar and isovector  effective masses were investigated by calculating charge and neutron distributions, 
single particle energies for a few orbits in $^{16}$O, $^{40}$Ca, $^{208}$Pb, and cross sections in $(e, e'p)$, 
and by comparing the result with experimental data.

In light nuclei $^{16}$O and $^{40}$Ca, the distributions of proton and neutron are insensitive to the effective masses, 
but in $^{208}$Pb, the effect of the effective mass appears clearly.
In particular, the difference of charge distribution at $r < 1$ fm is due to the isoscalar effective mass.
For the single particle energies, our theoretical results deviate at most about 18 \% from the experimental data.
From the energy levels, the small isoscalar effective masses describe the experimental values well in light nuclei, 
but for $^{208}$Pb the isoscalar effecitve masses that are close to 1 reproduce the experiment better than the light effective masses.

Reduced cross sections of exclusive $(e, e'p)$ reaction from $^{16}$O, $^{40}$Ca, and $^{208}$Pb agree well with experimental data.
In the light nuclei $^{16}$O and $^{40}$Ca, contribution of the effective mass is negligible.
For $^{208}$Pb, however, cross section depends clearly on the isoscalar effective mass: 
cross sections become large with smaller isoscalar effective mass.
Although the left-right asymmetry is insensitive to the effective mass, 
the contribution of the mass to $R_{LT}$ tends to be large with smaller mass.
Value of SF to reproduce $R_{LT}$ could be different from that for reduced cross section.
On the other hand, contribution of the isovector effective mass is vanishingly small.

Spectroscopic factors have been adjusted so that the KIDS0 model reproduces the cross section data.
We obtained SFs in the range 0.6--0.8 for light nuclei and 0.9--0.95 for $^{208}$Pb.
Our results are compatible with experiment and other theory.
We showed that SF could be dependent on the kinematic conditions such as energy and angle.

Response functions were calculated for $^{16}$O in the longitudinal-transverse channel, and compared with data from JLab.
Response function in the 1p$_{1/2}$ orbit depends on the effective mass sensitively, so even the difference between
$\mu_S=0.9$ and $\mu_S=1.0$ could be identified clearly.
Left-right asymmetry, on the other hand, does not show notable dependence on the effective mass,
and the theory result agrees well with data.

In conclusion, we have confirmed that quasi-electron scattering provides a useful tool
to study the effective mass of the nucleon in nuclear medium.


\section*{Acknowledgments}
This work was supported by the National Research Foundation of Korea (NRF) grant funded by the Korea govenment
(No. 2018R1A5A1025563 and No. 2020R1F1A1052495).

\end{document}